\documentclass[a4paper,11pt]{article}
\pdfoutput=1

\usepackage{mathtools}
\usepackage{cases}
\usepackage{booktabs}
\usepackage{jcappub}

\usepackage[T1,T2A]{fontenc}
\usepackage[utf8]{inputenc}
\usepackage[russian,english]{babel}

\usepackage{newunicodechar}

\newunicodechar{℈}{\azeriE}
\newunicodechar{ə}{\azerie}

\DeclareRobustCommand{\azeriE}{%
  {\fontencoding{T2A}\selectfont\symbol{"9A}}%
}
\DeclareRobustCommand{\azerie}{%
  {\fontencoding{T2A}\selectfont\symbol{"BA}}%
}

\makeatletter
\expandafter\def\expandafter\@uclclist\expandafter
  {\@uclclist\azerie\azeriE}
\makeatother

\definecolor{dred}{RGB}{163,35,47}
\definecolor{dblue}{RGB}{45,90,195}
\newcommand{\rar}{\rightarrow}

\newcommand{\dd}{\mathrm{d}}
\newcommand{\avg}[1]{\left\langle#1\right\rangle}

\newcommand{\al}{\alpha}
\newcommand{\be}{\beta}
\newcommand{\ga}{\gamma}
\newcommand{\de}{\delta}
\newcommand{\la}{\lambda}

\newcommand{\ep}{\epsilon}
\newcommand{\rh}{\rho}
\newcommand{\si}{\sigma}
\newcommand{\ta}{\theta}
\newcommand{\om}{\omega}
\newcommand{\Om}{\Omega}
\newcommand{\io}{\iota}
\newcommand{\ze}{\zeta}

\newcommand{\vka}{\varkappa}
\newcommand{\vep}{\varepsilon}
\newcommand{\vpi}{\varpi}

\newcommand{\vepT}{\vep_{\rm T}}
\newcommand{\vepV}{\vep_{\rm V}}
\newcommand{\vepS}{\vep_{\rm S}}

\newcommand{\denu}{{\de\om_b}}
\newcommand{\nnut}{{n \om_b t}}
\newcommand{\dPb}{\dot{P}_b}

\newcommand{\rhoDM}{\rho_{\text{DM}}}

\newcommand{\wm}{\frac{\om_b}{m}}

\newcommand{\deq}{\coloneqq}

\newcommand{\Eq}[1]{Eq.~(\ref{#1})}
\newcommand{\Eqs}[2]{Eqs.~(\ref{#1}) and~(\ref{#2})}
\newcommand{\Eqss}[2]{Eqs.~(\ref{#1})-(\ref{#2})}

\newcommand{\nn}{\nonumber} 

\newcommand{\M}{M}
\newcommand{\Mij}{\M_{ij}}
\newcommand{\mpl}{M_{\text{P}}}
\newcommand{\D}{\nabla}

\newcommand{\te}{\text{\itshape{ə}}}

\usepackage[small,labelfont=bf,textfont=it]{caption}
\usepackage{float}

\usepackage{newfloat}
\DeclareFloatingEnvironment[
    fileext=lop,
    listname={List of Panels},
    name=Panel,
    placement=tbhp,
    within=none,
]{panel}

\title{\textcolor{dblue}{Binary Pulsars as probes for Spin-2 Ultralight Dark Matter}}

\author[a]{Juan Manuel Armaleo,}
\author[a]{Diana L\'opez Nacir,}
\author[b]{and Federico R.~Urban}

\affiliation[a]{Departamento de F\'isica and IFIBA, FCEyN UBA, Facultad de Ciencias Exactas y Naturales\\Ciudad Universitaria, Pabellon I, 1428 Buenos Aires, Argentina}
\affiliation[b]{CEICO, Institute of Physics of the Czech Academy of Sciences\\Na Slovance 2, 182 21 Praha 8, Czech Republic}

\abstract{Binary pulsars can be excellent probes  of ultra-light dark matter.  We consider the scenario where the latter is represented by a spin-2 field.  The coherent oscillations of the dark matter field perturb the dynamics of binary systems, leading to secular effects for masses that resonate with the binary systems.  For the range \( 10^{-23} \)~eV~\(\lesssim m \lesssim 10^{-17} \)~eV we show that current timing data could potentially constrain the universal coupling strength of dark matter to ordinary matter at the level of \( \al \simeq 10^{-5} \).}

\begin{document}
\maketitle
\flushbottom


\section{Introduction}
\label{sec:introduction}

The modern cosmological model, known as \( \Lambda \)CDM, in order to reproduce observations, assumes that there exists a dark contribution to the total matter density of the Universe, dubbed Cold Dark Matter (CDM).  This component is typically associated to physics beyond the known fields of the Standard Model of particle physics (SM), consisting of particles that are moving very slowly (thence the ``Cold'' appellative) due to their heavy masses and very weak interactions with the thermal bath (Weakly Interacting Massive Particles or WIMPs), and thus characterised by non-relativistic phase-space distributions.

In this paper we focus on an alternative scenario, where Dark Matter (DM) is ultralight (ULDM) and described by a classical field.  The main assumptions are: I) in the early Universe the dynamics leads to a practically homogeneous background field on large scales; II) before matter domination era, the Hubble rate \( H \) becomes smaller than the mass of the field (hence \( m\gg H_{eq} \simeq 10^{-28} \)~eV, with \( H_{eq} \) the Hubble rate at epoch of matter-radiation equality), and after that the field oscillates with a frequency essentially given by its mass, with negligible self-interactions~\footnote{More precisely, the frequency can be approximated by \( m(1+V^2/2) \) with \( V \) being the effective velocity field of DM, which in the galactic halo can be estimated to be the virial velocity and is therefore very small: the mean Milky Way halo velocity is \( V_0\sim 10^{-3} \).  This means that the field remains coherent in frequency over at least \( 10^5 \) periods of oscillations\label{foot1}.}; III) the amplitude and phase of the field is determined by the DM density and velocity fields.

The traditional fields that follow this pattern are axion-like particles and dilatons~\cite{Preskill:1982cy,Abbott:1982af,Dine:1982ah,Turner:1983he,Hu:2000ke,Marsh:2015xka,Lee:2017qve}.  However, with these same assumptions, nearly the same late Universe cosmological evolution and phenomenology can be obtained from a massive vector or spin-2 tensor field, as shown respectively in~\cite{Nelson:2011sf,Cembranos:2012kk,Cembranos:2012ng,Cembranos:2013cba,Arias:2012az,Graham:2015rva,Knapen:2017xzo} and~\cite{Marzola:2017,Aoki:2017cnz}~\footnote{In the case of vectors and tensors, in addition to the coherence in frequency mentioned in footnote~\ref{foot1}, we also assume that the field remains coherent in direction over the same length and time scales.}.  Such different models also come equipped with a variety of possible interactions with the SM fields.  Two relevant questions are: 1) To what extent it is possible to discriminate between different masses, spins, and interactions?  2) Where can we look for the imprints of the nature and properties of these DM candidates?

Several observables have been already identified in the literature, and they allow us to probe different mass ranges and properties of ULDM~\cite{Irsic:2017yje,Armengaud:2017nkf,Zhang:2017chj,Bullock:2017xww,Bar:2018acw,Robles:2018fur,Baryakhtar:2017ngi,Baumann:2018vus,Marsh:2018zyw,Nebrin:2018vqt,Safarzadeh:2019sre,Wasserman:2019ttq}, precision timing measurements of binary pulsars offer a unique possibility to test the properties of such DM candidate, as was observed in~\cite{Blas:2016ddr}.  The robustness of the conclusions about these models clearly depends on the observable, the theoretical and experimental uncertainties, the assumptions made in  modelling, etc.  This highlights the importance of looking for alternatives either to probe other regions of the parameter space or to test the same region in an independent way.  On general grounds, one expects characteristic signatures in the phenomenology of the different candidates to be relevant when spatial gradients and pressure effects become important (i.e., when the description as a perfect fluid with no pressure breaks down due to inhomogeneities).  Alternatively, another regime is when the time scale of oscillations of the ULDM field becomes comparable to that of the evolution of the observable under study, potentially leading to resonant, secular effects.  In this paper we study this second regime.

We consider the possibility that ULDM is given by a spin-2 field, and we focus on the mass range \( 10^{-23} \)~eV\( \lesssim m \lesssim 10^{-17} \)~eV.  Following previous studies for the scalar and vector ULDM models~\cite{Blas:2019hxz,LopezNacir:2018epg}, here we look at the very precise measurements of the orbital parameters of binary pulsars to probe the interactions between the ULDM spin-2 field and the ordinary matter of which the stars are made.

In Sec.~\ref{sec:action} we introduce the spin-2 field and its properties; we will base our discussion on~\cite{Marzola:2017} as a practical blueprint for our discussion, but our results are not limited to that specific model.  In the case of universal coupling the interactions between the ULDM and ordinary matter are given in terms of a single parameter, \( \al \).  Currently, the best constrains on \( \al \) for the mass range we consider are given by~\cite{Hohmann:2017uxe,Sereno:2006mw}.

In Sec.~\ref{sec:secular}, we show that due to the ULDM-SM interaction the coherent oscillations of the field affect the dynamics of binary systems, leading to secular effects.  Sec.~\ref{sec:pheno} is dedicated to the phenomenology.  There we show that for several masses in the range we are considering, the constraints we can obtain with this method are comparable or even better than those previously obtained in the literature.  In Sec.~\ref{sec:end} we summarise our conclusions.

\section{Spin-2 ULDM}
\label{sec:action}

\subsection{The spin-2 field}\label{ssec:spin2}

A massive spin-2 field \( \M_{\mu\nu} \) is described by the Fierz-Pauli lagrangian density
\begin{align}
  {\cal L} &\, \deq \frac12 \M_{\mu\nu}\mathcal{E}^{\mu\nu\rh\si}\M_{\rh\si} - \frac14 m^2 \left( \M_{\mu\nu}\M^{\mu\nu} - \M^2 \right) \,, \label{eq:lagrangian}
\end{align}
where \( \M \deq g^{\mu\nu} \M_{\mu\nu} \), and the Lichnerowicz operator \( \mathcal{E}^{\mu\nu\rh\si} \) is defined by
\begin{align}
  \mathcal{E}^{\mu\nu}_{~~\rh\si} \deq &\, \de^\mu_\rh \de^\nu_\si \Box - g^{\mu\nu} g_{\rh\si} \Box + g^{\mu\nu} \D_\rh \D_\si + \nn \\
  &\, + g_{\rh\si} \D^\mu \D^\nu - \de^\mu_\si \D^\nu \D_\rh - \de^\mu_\rh \D^\nu \D_\si \,. \label{eq:lich}
\end{align}
This lagrangian arises for example in bimetric theory, where two spin-2 fields coexist in four dimensions, in the limit where \( m\gg  H \), see~\cite{Marzola:2017,Aoki:2017cnz}.  In this case the theory can be seen as a massive spin-2 field in a standard Friedman-Lema\^itre-Robertson-Walker (FLRW) cosmology described by the metric \( g_{\mu\nu} \).

The Bianchi identities ensure that the massive field \( \M_{\mu\nu} \) propagates only five degrees of freedom, which, for a homogeneous and isotropic FLRW background can be conveniently chosen to be the six \( \Mij \) components, subject to the additional tracelessness constraint \( \M^i_{~i}=0 \).  The equations of motion for the ULDM field in the late Universe then read
\begin{align}
  \ddot\M_{ij} + 3H\dot\M_{ij} -\bigtriangleup \Mij+ m^2\Mij &\, = 0 \,. \label{eq:eom}
\end{align}
This equation is reminiscent of that for scalar ULDM and it has the same solutions. The homogeneous background solution is given by
\begin{align} 
  \Mij &\, = \frac{\hat{\M}_{ij}}{ R^{3/2}} \cos{(mt+\Upsilon)}\vep_{ij} = \frac{\sqrt{2\rhoDM}}{m R^{3/2}}\cos{(mt+\Upsilon)}\vep_{ij} \,, \label{eq:hij}
\end{align}
where \( R \) is the scale factor of the Universe, the overall amplitude has been fixed so that the ULDM energy density matches the observed \( \rhoDM \), \( \Upsilon \) is a random phase, and \( \vep_{ij} \) is an angular quadrupole matrix with unit norm, zero trace and is symmetric, see Appendix~\ref{app:angular}.  With this solution the ULDM energy density \( \rhoDM \sim R^{-3} \), and the pressure \( P_\text{DM} \) averages to zero on the large time-scales relevant for the cosmological background evolution.  

 One interesting diversion on bimetric theory is partly-massless gravity~\cite{Deser:2001us,Hassan:2012gz}.  Without going into details, the crucial feature of that model is that, under some specific conditions, only the \emph{four} degrees of freedom of the \( \Mij \) associated with the spin-1 and spin-2 polarisations do propagate, whereas the spin-0 polarisation does not.  Our results also apply to this case as it will be clear below.

Another possibility one could think of in this context is higher-dimensional spin-2 fields; once the reduction (or compactification) to four dimensions is performed, the higher-dimensional fields turn into a four-dimensional tower of massive states, with masses set by the parameters of the compactification scheme~/~mechanism~\cite{Aulakh:1985un,Bonifacio:2016blz}.  In this case the coupling to ordinary matter may be non-universal; however our results can be generalised to this case.

On  scales relevant for binary pulsars, the  local ULDM field can be written as in  Eq.~(\ref{eq:hij}) where   the density $\rhoDM$ and phase $\Upsilon$ are now given by   their local  values, which
will  depend on the spatial location of the binary inside the ULDM halo. As for the scalar case, one expects gradients of the field to be relevant at scales of order of the de Broglie wavelength  $\lambda_{\rm dB}\equiv (m V)^{-1}$, where $V$ is the effective velocity of the ULDM.  In what follows we   will work at leading order in the post-Newtonian expansion. Therefore, assuming gradients of order $\lambda_{\rm dB}^{-1}$, we will keep only the leading order in the gradients and neglect higher order derivatives of the ULDM field.  

\subsection{DM interactions with the stars}\label{ssec:stars}

The interaction between the spin-2 ULDM and ordinary matter is given by the action
\begin{align}\label{eq:lagrdens}
  S_\text{int} &\, \deq \la \int\dd^4x\,\sqrt{-g} \M_{\mu\nu} T^{\mu\nu} \,,
\end{align}
where \( \la \deq \al/2\mpl \) is the interaction strength~\footnote{In bimetric theory the parameter \( \al \) has three roles: it is the interaction strength of the massive field with the SM, it is the mixing parameter between the massless and massive eigenstates of the theory, and it parametrises the strength of the self-interactions of one metric with respect to the other.  In the limit of \( \al\rar0 \) the theory reduces to General Relativity.}, \( \mpl\approx2.4\times10^{18} \)~GeV is the reduced Planck mass, and \( T_{\mu\nu} \) is the energy momentum tensor (EMT) of standard matter.

For our purposes we can approximate each star in a binary system, labelled as 1 and 2, as a point particle with mass \( M_A \) (with \( A=[1,2] \)), energy \( E_A \), position \( \vec{x}_A \), and 4-velocity \( u^\mu_A \); the EMT for the system would then be
\begin{align}
  T^{\mu\nu} &\, = E_1 u_1^\mu u_1^\nu \de(\vec{x}-\vec{x}_1) + E_2 u_2^\mu u_2^\nu \de(\vec{x}-\vec{x}_2) \,. \label{eq:EMT}
\end{align}

Since the stars are non-relativistic \( E_A \simeq M_A \) and \( u^i_A = v_A^i \), with \( v_A^i \deq \dd x_A^i / \dd t \) the (non-relativistic) velocity of the body.  Moreover, since \( \partial_{\mu}\M^{\mu\nu}=0 \), one can see that $M_{0i}$ ($M_{00}$) is of first (second) order in gradients of $M_{ij}$. Therefore, the interaction lagrangian at leading order is given by
\begin{align}
  L_\text{int} &\, = \la M_T  \left[ M_{00} +2M_{0i}V_{\rm CM}^i+ \Mij V_{\rm CM}^i V_{\rm CM}^j\right] + \la    \mu v^i v^j    \Mij \,, \label{eq:lagr}
\end{align} where $M_T$ is the total mass is \( M_T \deq M_1 + M_2 \), \(\mu\deq M_1 M_2/M_T \) ,  \( v^i \deq v_1^i - v_2^i \)  the relative velocity of the stars, and \( V_{\rm CM}^i \deq  (M_1 v_1^i+ M_2 v_2^i)/M_T\) the  center of mass velocity of the binary system.

\section{Secular effects}
\label{sec:secular}

In this section we use the method of osculating orbits to compute the secular effects on the orbital parameters.  We start by noticing that the equations for the  center of mass  of the binary system decouple from the ones  for  \( r^i \deq r_1^i - r_2^i \) that describe the relative motion.  Moreover, the perturbation  on the center of mass oscillates as the ULDM field and averages to zero over  time scales much longer than the  period of oscillation  and, therefore, it does not produce a secular effect. On the other hand, we can express the equation of motion for the orbital motion as
\begin{align}
  \dot{v}_i + 2\la \left(\Mij \dot{v}^j + \dot{\M}_{ij} v^j\right) + \frac{G M_T}{r^3} r_i &\, = 0 \,, \label{eq:eom_v}
\end{align}
where \( G\deq8\pi/\mpl^2 \) is Newton's constant.  Using the fact that the unperturbed orbit is described by \( \dot{v}_i = - G M_T r_i / r^3 \), we can derive the expression for the perturbation, parametrised in terms of a force per unit mass \( F_i \), to the relative acceleration between bodies as \( \dot{v}_i \rar \dot{v}_i + \de\dot{v}_i \deq \dot{v}_i + F_i \):
\begin{align}
  F_i &\, = 2\la \left[ \frac{G M_T}{r^3} \Mij r^j - \dot{\M}_{ij} v^j \right] \,. \label{eq:force_hij}
\end{align}
Using now \Eq{eq:hij}, and taking into account that \( r^j = r\hat{r}^j \) and \( v^j = \dot{r}\hat{r}^j + r\dot{\ta}\hat{\ta}^j \), where we choose the reference frame of the binary system with polar coordinates \( (r,\ta,z) \) (the  unit vectors are denoted as \( (\hat{r},\hat{\ta},\hat{z}) \)), \( F_i \) can be recast as
\begin{align}
  F_i = &\, \sqrt2\vka \left\{ \wm\left( \frac{a}{r} \right)^2 \hat{r}^j \cos(mt+\Upsilon) \right. \nn \\
  &\, \left. + \frac{1}{\te} \left[ \hat{r}^j e\sin\ta + \hat{\ta}^j \left(1+e\cos\ta\right) \right] \sin(mt+\Upsilon) \right\} \vep_{ij} \,, \label{eq:fi}
\end{align}
where \( \vka \deq 2\la a \om_b \sqrt{\rhoDM} \),
\begin{align}
  P_b & \deq \frac{2\pi}{\om_b} = 2\pi\sqrt{\frac{a^3}{GM_T}} \,,
\end{align}
is the orbital period, \( \om_b \) the orbital frequency, \( a \) is the semi-major axis of the system, \( e \) its eccentricity, and we defined \( \te \deq \sqrt{1-e^2} \) for brevity.

The perturbations caused by the ULDM on the binary system oscillates; however, as noted in~\cite{Blas:2016ddr,Blas:2019hxz,LopezNacir:2018epg}, if the  ULDM oscillation frequency \( m \) is close to  the orbital frequency \( \om_b \)   or  an integer multiple of it  \( N\om_b \), with $N \in \mathbb{N}$, the system will experience secular effects due to the resonant behaviour of the perturbations.

\begin{figure}[tbhp]
  \center{\includegraphics[width=8cm]{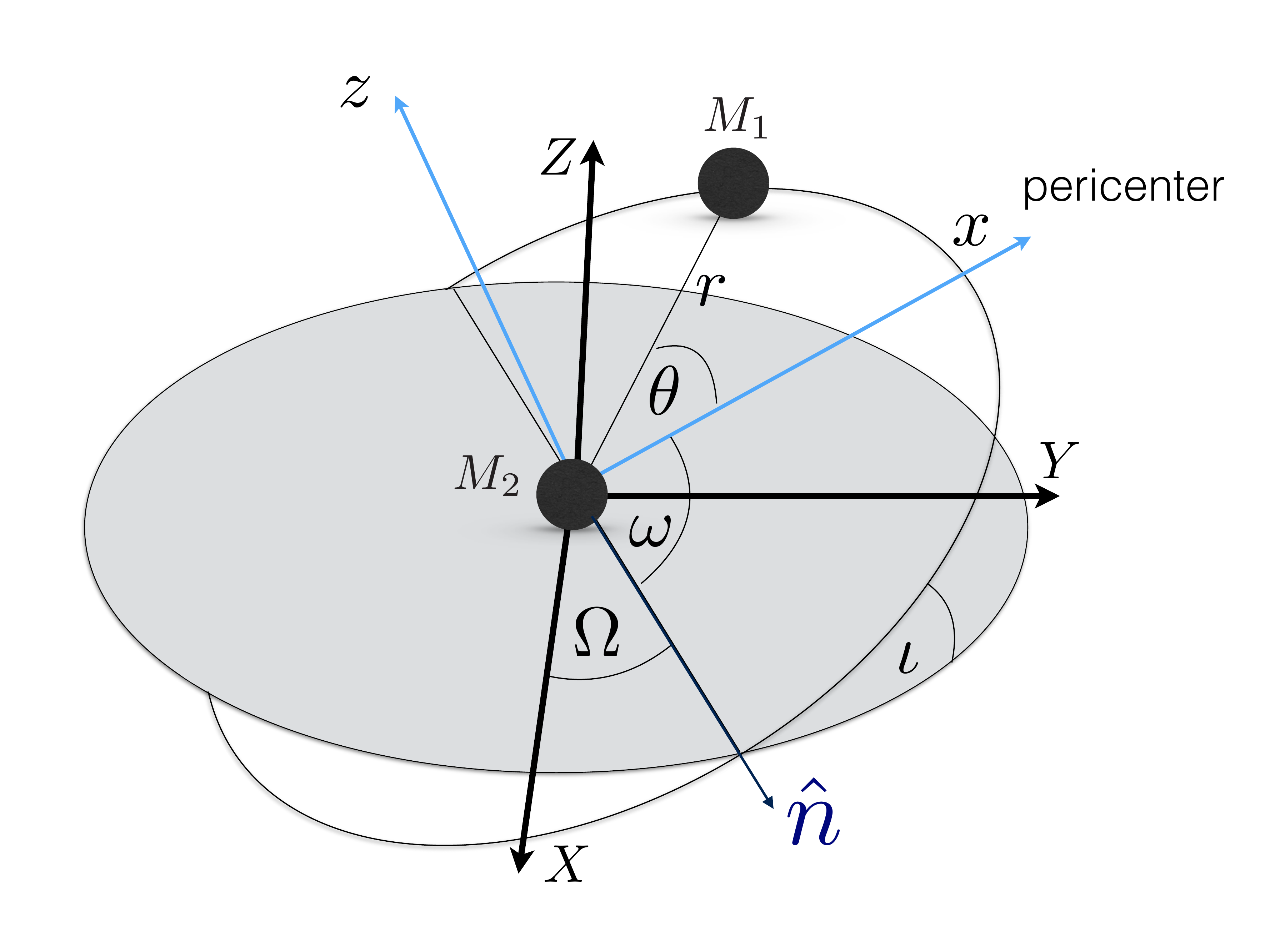}}
  \caption{Description of Keplerian orbits in terms of the orbital elements viewed in the fundamental reference frame \( (X,Y,Z) \).  The cartesian orbital frame \( (x,y,z) \) and the polar one \( (r,\ta,z) \) are also shown (centered at \( M_2 \) for convenience).}\label{fig:orbits}
\end{figure}

In order to obtain the secular contribution to the variation of each orbital parameter we proceed as in~\cite{Blas:2019hxz,LopezNacir:2018epg}, starting from the Lagrange Planetary equations (see Appendix~\ref{app:bessel}).  To make the paper as self-contained as possible, here we summarise the procedure for the semi-major axis \( a \), while we provide the equations for all other orbital parameters in Appendix~\ref{app:bessel}.  In the Post-Keplerian formalism of osculating orbits~\cite{Danby:1970} (see also~\cite{Blas:2019hxz,LopezNacir:2018epg}) the time derivative of \( a \) is given by
\begin{align}
  \frac{\dot{a}}{a}	= &\, \frac{2}{\om_b}\left\{\frac{e\sin\ta}{a\te} F_r + \frac{\te}{r} F_\ta \right\} \,, \label{eq:a}
\end{align}
where we have decomposed the vector \( F^i \) in the reference frame of the binary system as \( \vec{F}=F_r\hat{r}+F_{\ta}\hat{\ta}+F_z\hat{z} \) (see figure~\ref{fig:orbits}). 

The polarisation tensor $\vep_{ij}$ (defined in the same cartesian orbital frame) can be decomposed into five independent spherical tensors with real coefficients (see Appendix~\ref{app:angular} for details):
\begin{align}
  \vep_{ij} & = \frac{1}{\sqrt2}\left(
  \begin{array}{ccc}
    \vepT c_{\chi} - \vepS/\sqrt3 & \vepT s_{\chi} & \vepV c_{\eta} \\
    \vepT s_{\chi} & - \vepT c_{\chi} - \vepS/\sqrt3 & \vepV s_{\eta} \\
    \vepV c_{\eta} & \vepV s_{\eta} & 2\vepS/\sqrt3 \\
  \end{array} \right) \,, \label{eq:pol}
\end{align}
where  we have defined  three real parameters,  $\vepS$, $\vepV$ and $\vepT$, satisfying 
\( {\vepS}^2+{\vepV}^2+{\vepT}^2=1 \), and  two angular variables,  $\eta$ and $\chi$. Here and in what follows we employ the shortcut notation \( s_x \deq \sin{x} \), \( c_x \deq \cos{x} \).
The above parametrization defines a  scalar, a vector, and a tensor component of $\vep_{ij}$, respectively,  as the contribution proportional to  $\vepS$, $\vepV$ and $\vepT$
~\footnote{The separation in scalar, vector, and tensor components depends on the choice of reference frame. Our definition is convenient for  understanding and visualising the interplay between the five components of the oscillating quadrupole and the binary system, and to make connection with the previously studied cases of a scalar perturbation.}.

Using \Eq{eq:pol} we can write explicitly:
\begin{subequations}\label{eq:forces_comp}
  \begin{align}
    F_r & = \vka \left\{ \wm \left(\frac{a}{r}\right)^2 \left[ \vepT c_{\chi-2\ta} - \frac{\vepS}{\sqrt3} \right] c_{mt+\Upsilon} + \frac{1}{\te} \left[ \vepT s_{\chi-2\ta} + e \vepT s_{\chi-\ta} - e\frac{\vepS}{\sqrt3} s_\ta \right] s_{mt+\Upsilon} \right\} \,, \label{eq:fr} \\
    F_\ta & = \vka \left\{ \wm \left(\frac{a}{r}\right)^2 \left[ \vepT s_{\chi-2\ta} \right] c_{mt+\Upsilon} - \frac{1}{\te} \left[ \vepT c_{\chi-2\ta} + e \vepT c_{\chi-\ta} + \frac{\vepS}{\sqrt3}(1+e c_\ta) \right] s_{mt+\Upsilon} \right\} \,, \label{eq:fta} \\
    F_z & = \vka \left\{ \wm \left(\frac{a}{r}\right)^2 \left[ \vepV c_{\eta-\ta} \right] c_{mt+\Upsilon} + \frac{1}{\te} \left[ \vepV s_{\eta-\ta} + e \vepV s_{\eta} \right] s_{mt+\Upsilon} \right\} \,. \label{eq:fz}
  \end{align}
\end{subequations}

As detailed in~\cite{Blas:2019hxz,LopezNacir:2018epg}, we obtain secular variations of the orbital parameters when the binary system is in resonance with the oscillating background perturbations, as these oscillations enter directly in the expressions for the force \Eqss{eq:fr}{eq:fz}.  Expressing the orbit in terms of Bessel series in \( \sin[n\om_b(t-t_0)] \) and \( \cos[n\om_b(t-t_0)] \), with \( t_0 \) the time of periastron, and parameterising the (small) gap between the two frequencies as \( \denu \deq m - N\om_b \), where \( N \) is the resonance harmonic number and \( \denu \ll m \), then, upon averaging over a long time \( \Delta t \) for which \( P_b /N\ll \Delta t \ll 2\pi/\denu \), we have
\begin{align}
  \avg{s_{ n \om_b (t-t_0)} s_{m t+\Upsilon}} \approx \, \frac12  c_{\ga(t)} \de_{n,N}  \,, & \quad
  \avg{s_{ n \om_b (t-t_0)} c_{m t+\Upsilon}} \approx \, -\frac12 s_{\ga(t)}  \de_{n,N}  \,, \nn\\
  \avg{c_{ n\om_b (t-t_0)} s_{ m t+\Upsilon}} \approx \, \frac12  s_{\ga(t)} \de_{n,N}  \,, & \quad
  \avg{c_{ n\om_b (t-t_0)} c_{ m t+\Upsilon}} \approx \, \frac12 c_{ \ga(t)}\de_{n,N}   \,,
\end{align}
with
\begin{align}
  \gamma(t) \deq \denu(t-t_0)+mt_0+\Upsilon \,.
\end{align}

Keeping only the dominant secular term \(n=N\) and using \( \dPb / P_b = 3\dot{a}/2a \), we obtain:
\begin{align}
  \avg{\dPb} = -2\la P_b \sqrt{3\rhoDM} \left\{ \vepS J_N(Ne) s_{\ga(t)} + \vepT \left[ \mathcal{F}_+(N,e) s_{\ga(t)+\chi} + \mathcal{F}_-(N,e) s_{\ga(t)-\chi} \right] \right\} \,, \label{eq:dPb}
\end{align}
with \(J_N(z)\) the Bessel functions of the first kind and
\begin{subequations}
  \begin{align}
    \mathcal{F}_+(N,e) & \deq \frac{\sqrt3}{4} \left[2J_N(Ne) +  \frac{2eJ'_N(Ne)}{\te} + \left( \frac{\tilde{B}_N(Ne)-{B}_N(Ne)}{\te^2} \right)\right] \,, \label{eq:defFp} \\
    \mathcal{F}_-(N,e) & \deq \frac{\sqrt3}{4} \left[2J_N(Ne) -  \frac{2eJ'_N(Ne)}{\te} - \left( \frac{\tilde{B}_N(Ne)+{B}_N(Ne)}{\te^2} \right) \right] \,. \label{eq:defFm}
  \end{align}
\end{subequations}
The coefficients \( \tilde{B}_N(Ne) \) and \( {B}_N(Ne) \) are given in Appendix~\ref{app:bessel} in terms of  Bessel functions.  The choice of factorisation in the \( \mathcal{F}_+(N,e) \) and \( \mathcal{F}_-(N,e) \) functions comes from the fact that they will behave very differently in the limit of circular orbits, as we will see in Sec.~\ref{sec:pheno}.

\section{Phenomenology}
\label{sec:pheno}

\subsection{Orbital period}\label{ssec:orbP}

We start by analysing the secular effect of the DM field on the orbital period.  From \Eq{eq:dPb} we see that only the tensor and scalar components of the spin-2 field contribute to the drift in the orbital period.

In order to place a constraint we took the observed value for the orbital period change, and, where available, we subtracted all kinematic effects (Shklovskii effect, differential Galactic rotation, Galactic potential), and the change due to gravitational wave damping; in one case (B1259-63) we subtracted also the effect due to the mass loss from the companion due to its stellar wind.  If not available, we used the upper bound estimated in the corresponding references, which we provide in Table~\ref{tab:frac}.  We call this the central value of the ``intrinsic'' secular period drift, which, absent any other effects, should be compatible with zero.

In order to be fully conservative, we then impose that the effect due to the oscillating ULDM be smaller than (the absolute value of) the central value of the ``intrinsic'' orbital period change plus its the error (obtained by adding all errors in quadrature); in this way we account for the largest possible deviation from zero as acceptable by the measurements, see also the discussion in~\cite{Blas:2019hxz}.  Notice that we have used the same criterium for the time-variation of the eccentricity.   We collect all the binary systems used in this analysis in Table~\ref{tab:frac}.

\begin{figure}[tbhp]
  \center{\includegraphics[width=0.8\textwidth]{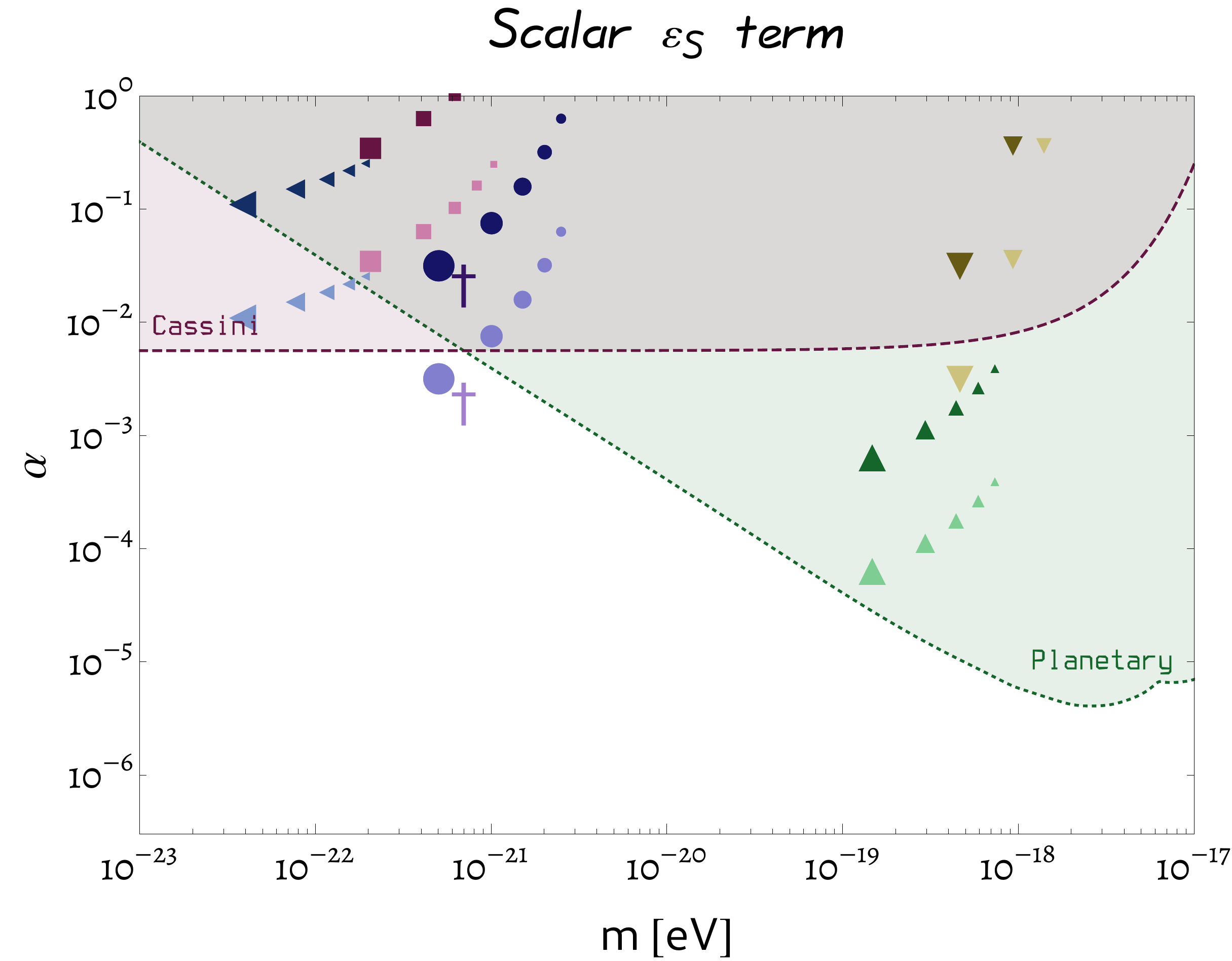}}\\
  \center{\includegraphics[width=0.8\textwidth]{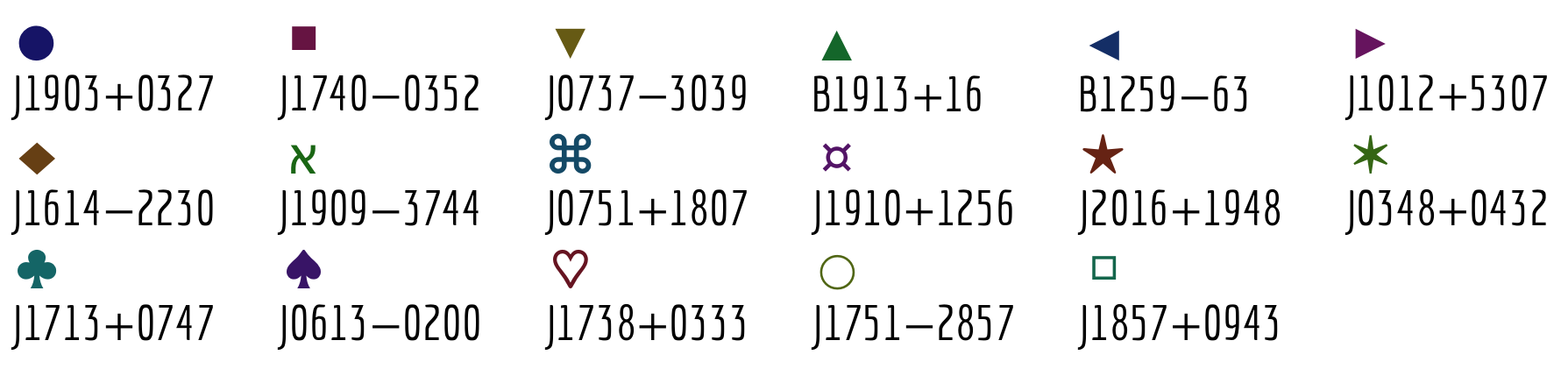}}\\\vspace{-4pt}
  \center{\includegraphics[width=0.45\textwidth]{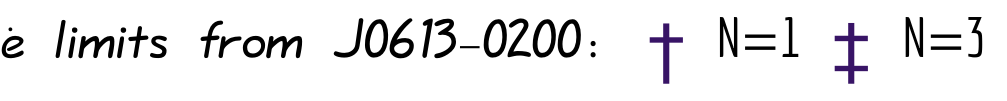}}
  \caption{Limits on the ULDM coupling \( \al \) versus the ULDM mass \( m \) obtained from the scalar polarisation \( \vepS \).  Dark coloured symbols are the current bounds obtained from the corresponding systems with parameters given in Table~\ref{tab:frac}.  The same symbols in lighter colours show the constraints that would be obtained for the same systems were the precision on \( \dPb \) a factor of 10 higher.  The largest symbols refer to the first resonance \( N=1 \), and the constraints for higher resonances (up to \( N=5 \)) are shown with the same symbols but progressively smaller sizes.  Symbols in the legend which do not appear in this figure are relevant for Figs.~\ref{fig:sin_plus} and~\ref{fig:sin_minus}; we collect all the binary systems used in this analysis in Table~\ref{tab:frac}.  The shaded region above the dashed purple line is excluded by solar system tests~\cite{Hohmann:2017uxe}.  The shaded region above the dotted green line is excluded by planetary constraints~\cite{Sereno:2006mw}.}\label{fig:mono}
\end{figure}

If we keep only \( \vepS \), the polarisation matrix \( \vep_{ij} \) becomes diagonal and, by construction, the eigenvalues of \( \vep_{ij} \) associated to the orbital plane are identical.  Therefore, as for the scalar case studied previously in~\cite{Blas:2016ddr,Blas:2019hxz} and as expected from  symmetry, the effect will survive only for eccentric orbits \footnote{Notice that the effective coupling of ULDM and the stars we are considering here differs from the universally coupling scalar interaction assumed in~\cite{Blas:2016ddr}.}.  We show the contraints obtained with current data in figure~\ref{fig:mono}.  Here and in what follows we assume a conservative value for the local DM density, \(\rhoDM=0.3 \,{\rm GeV}/{\rm cm}^3\).

The existing constraints on the coupling \( \al \) come from the Cassini tracking experiment~\cite{Hohmann:2017uxe}.  The planetary constraints are obtained by measuring the extra-precession of the planets of the inner solar system, see~\cite{Sereno:2006mw}; these constraints update and supersede those reported in~\cite{Adelberger:2009zz,Murata:2014nra} in part of the mass range.

In general, without tuning the angular parameters of the quadrupole, all the components of \( \vep_{ij} \) will contribute in a similar way and, as we show next, unlike the scalar case, also systems with circular orbits will be affected.  Indeed, the tensor component \( \vepT \) leads to a much richer phenomenology.  The effect on \( \dPb \) depends on the orientation of the tensor polarisations, as we can see from \Eq{eq:dPb}.  In particular, for circular orbits the \( N=2 \) harmonic contributes to the secular drift of the orbital period since \( \mathcal{F}_+(N,e) \rar \sqrt3\de_{N,2}/4 \) while \( \mathcal{F}_-(N,e) \rar e J'_N(Ne) \rar J_N(Ne) \rar 0 \) when \( e\rar0\) (see \Eq{Circularlimit}), whence
\begin{align}
  \avg{\dPb} \to - \frac32 \la P_b \sqrt{\rhoDM} \vepT s_{\ga(t)+\chi} \,. \label{eq:dPb_circ}
\end{align}

\begin{figure}[tbhp]
  \center{\includegraphics[width=0.8\textwidth]{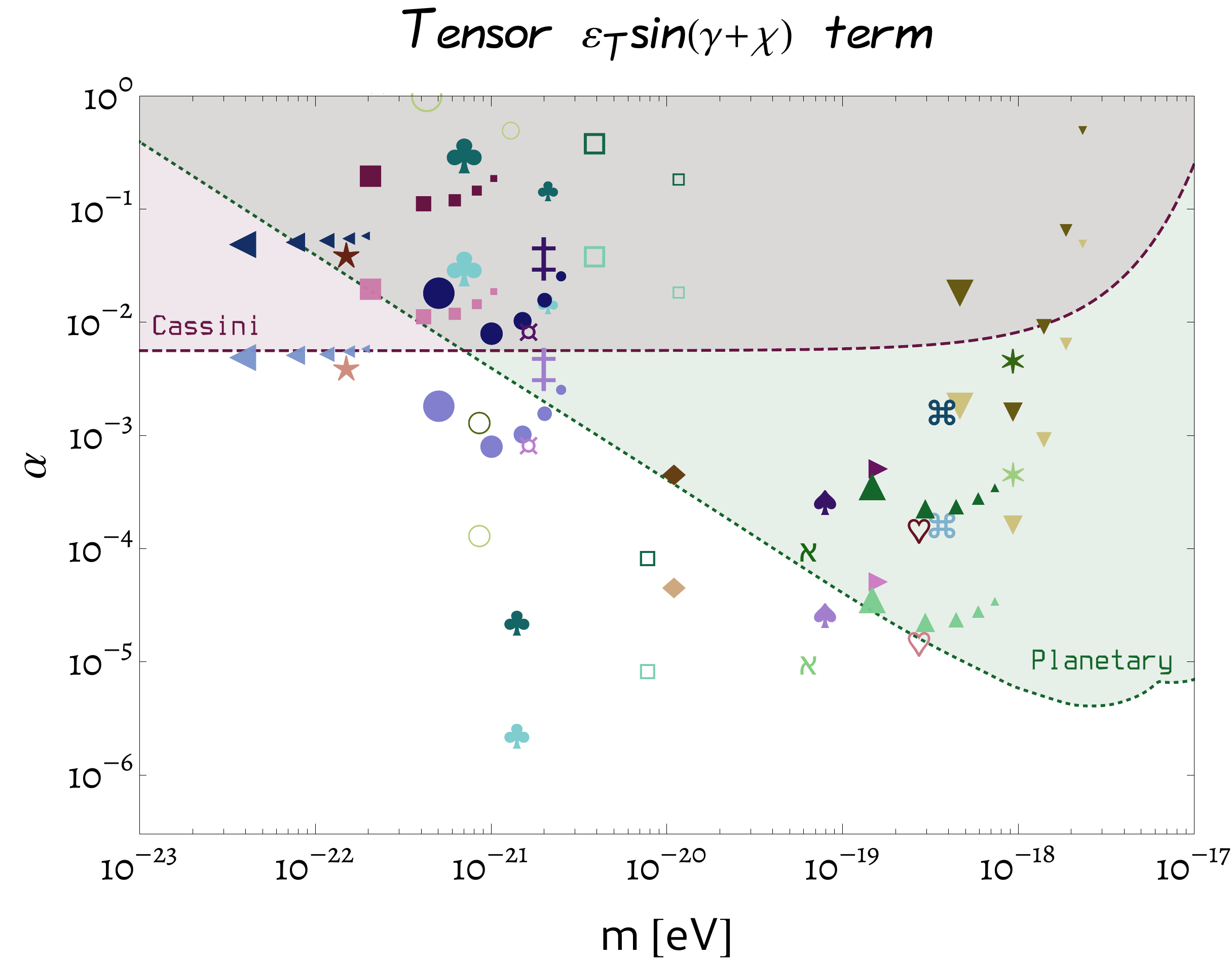}}
  \caption{Limits on the ULDM coupling \( \al \) versus the ULDM mass \( m \) obtained from the \( \mathcal{F}_+(N,e) \) contribution of the tensor polarisation \( \vepT \), see~\Eq{eq:dPb}.  See figure~\ref{fig:mono} for symbol references.}\label{fig:sin_plus}
\end{figure}

\begin{figure}[tbhp]
  \center{\includegraphics[width=0.8\textwidth]{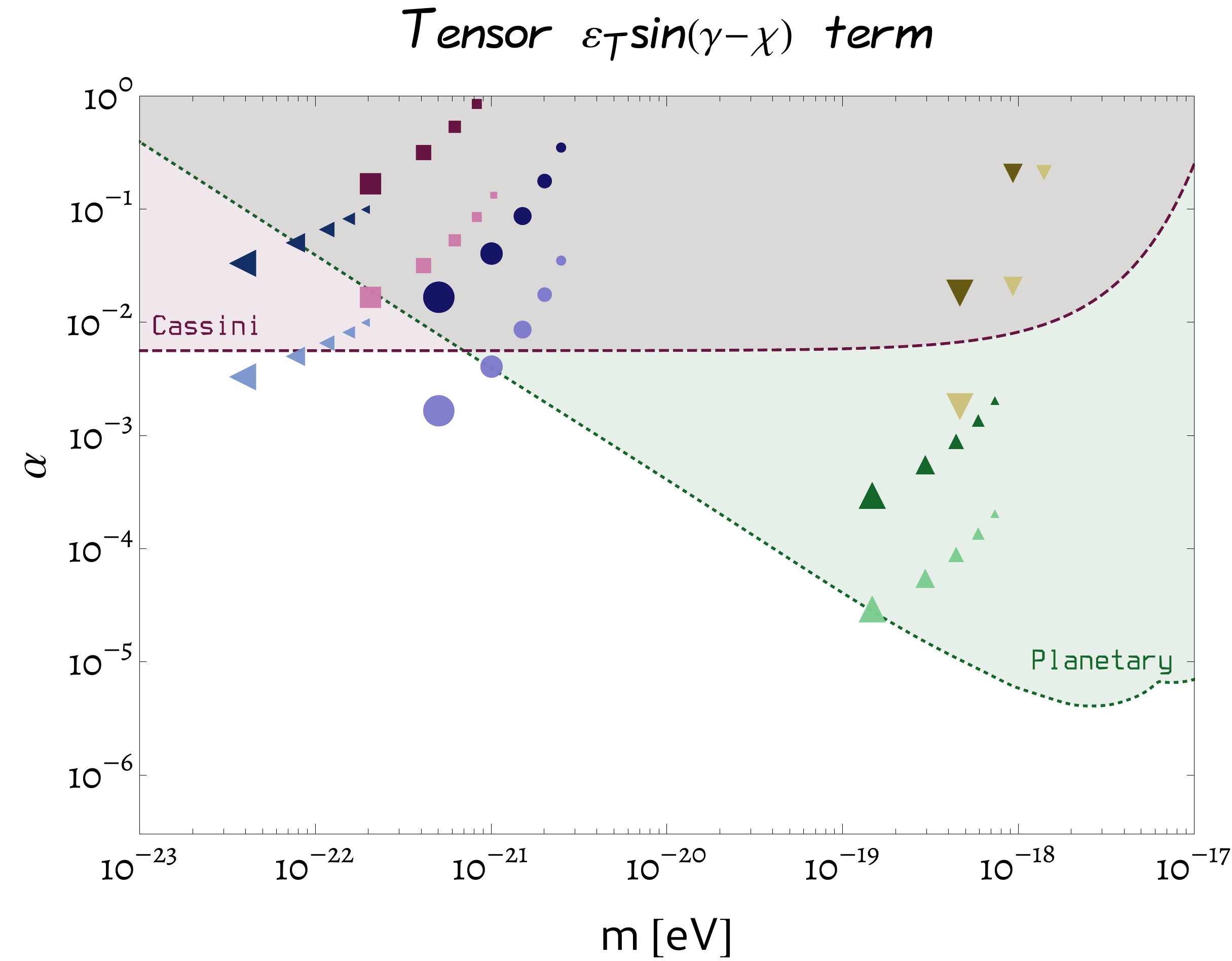}}
  \caption{Limits on the ULDM coupling \( \al \) versus the ULDM mass \( m \) obtained from the \( \mathcal{F}_-(N,e) \) contribution of the tensor polarisation \( \vepT \), see~\Eq{eq:dPb}.  See Fig.~\ref{fig:mono} for symbol references.}\label{fig:sin_minus}
\end{figure}

We show the limits on \( \al \) obtained from the two tensor polariations separately in figure~\ref{fig:sin_plus} for the \( \mathcal{F}_+(N,e) \) piece, and in figure~\ref{fig:sin_minus} for the \( \mathcal{F}_-(N,e) \) one.  It is worth emphasising that in figure~\ref{fig:sin_plus} all systems contribute, since this term does not vanish for circular orbits.  For example, we can observe the effect of the \( e\to0 \) limit in J0737-3039, which has \( e\approx0.1 \): the \( N=1 \) and \( N=[3,4,5] \) harmonics are significantly less constraining than the dominant \( N=2 \) contribution; as \( e\to0 \) this becomes more and more pronounced.  This is relevant because systems with near-circular orbits are much more common in nature than highly eccentric systems~\cite{Lorimer:2004,Lorimer:2008se}.  The \( \mathcal{F}_-(N,e) \) function instead vanishes when \( e\to0 \), so the only systems that contribute to figure~\ref{fig:sin_minus} are again the same ones that do in the scalar case.

Lastly, as we can see from \Eq{eq:dPb} the vector component \( \vepV \) does not contribute to the secular drift in orbital period.  This component generates a perturbation in the \( \hat{z} \) direction and will have an effect on the binary through the non-vanishing \( F_z \), see below for the discussion of the other orbital parameters.

\subsection{Other orbital parameters: nearly circular orbits}\label{ssec:orb_else}

So far we have focussed on the secular drift of the orbital period, \( \dPb \).  Analogously, all other orbital parameters might be affected secularly.  In the previous section we assessed the constraints on \( \al \) that can be achieved from the measurements of \( \dPb \), independently of the effect on the other orbital parameters.  The same analysis can be done independently for the drift of each orbital parameter.  Furthermore, one can expect to improve the constrains on \( \al \) by fitting data taking into account that the perturbation of the orbital parameters are not independent.  We provide all the relevant equations necessary for a quantitative and comprehensive analysis in Appendix~\ref{app:bessel}.  However, we leave such analysis for future work.  In this section, we briefly discuss on the effects on the other orbital parameters and, for simplicity, we consider only systems with near circular orbits.  From Panel~\ref{orb.param} in Appendix~\ref{app:bessel}, and the use of \Eq{Circularlimit}, it is immediate to obtain the secular changes of all six orbital parameters for \( e\to0 \).  In this limit, only three resonances give a non-vanishing contribution: those with \( N=1,2,3 \) that correspond to masses \( m\simeq\om_b \), \( m\simeq2\om_b \), and \( m\simeq3\om_b \).

The \( N=1,3 \) resonances yield a qualitatively different phenomenology from that with \( N=2 \).  As we have discussed in the previous section, only the \( N=2 \) resonance affects the orbital period in this limit.  Taking into account that also the other parameters are affected, we obtain the result (\( N=2 \) and \( e\to0 \)):
\begin{subequations}
  \begin{align}
    \avg{\dPb} = &\, -\frac{3\la}{2}{{P}_b} \sqrt{\rhoDM}\vepT s_{\ga(t)+\chi}\,, \\ 
           \avg{\dot e}		= & \,0\,,  \\
        \avg{\dot\Om} = &\, \frac{\la\sqrt{\rhoDM}\vepV}{4s_\io}s_{\ga(t)+\eta-\om}\, ,\\
    \avg{\dot\io}	           = &-\, \frac{\la\sqrt{\rhoDM}\vepV}{4}c_{\ga(t)+\eta-\om}\,, \\
       e\avg{\dot\vpi}		= & \,0\,,  \\
    \avg{\dot\ep_1} = &\,\la\sqrt{\rhoDM}\vepT c_{\ga(t)+\chi} + 2s^2_{\io/2}\avg{\dot{\Om}}\,. 
  \end{align}
\end{subequations}where for the argument of the periastron $\omega=\vpi-\Om$  the leading order contribution  goes as $1/e$ and vanishes in this case \footnote{The reason for keeping up to this order here is that for low-eccentricity orbits, the motion is more appropriately  parameterised in terms of the parameters $\eta= e \sin{\omega}$ and $\kappa=e\cos\omega$, which are the Laplace-Lagrange parameters that are actually used in the data analysis (see for instance \cite{Edwards:2006zg}).}.
Notice that also the vector component \( \vepV \) contributes, which (since this yields a perturbation that is ortogonal to the orbital plane) only produces an effect in the orientation of the orbit with respect to the fundamental  reference frame.  It is however too difficult to obtain a bound on this kind of effect from current measurements (see for instance~\cite{Damour:1991rd}).  The strongest constraints come from the measurements of \( \dPb \) that we have presented above.

For \( N=1 \) and \( N=3 \) the effect is similar to the Damour-Sch{\"a}fer effect~\cite{Damour:1991rq}, which has also been discussed in the context of ULDM models in~\cite{LopezNacir:2018epg,Blas:2019hxz} for other couplings to scalar and vector ULDM fields, and the equations can be recasted as $(e\to0)$:
\begin{align}\label{Standardwetozero}
  \avg{\dot{e}} = &\, \frac32 \frac{F^{\rm SEP, eff}_y}{a\om_b} \,\,,\,\,\,\, \quad \avg{\dot{\om}} = -\frac32 \frac{F^{\rm SEP, eff}_x}{e a\om_b} \,,\\
\avg{\dPb}= &        \avg{\dot\Om} =    \avg{\dot\io}	 =    \avg{\dot\ep_1} =0 \,, \end{align}
where, for \( N=1 \)  
\begin{align}
  \vec{F}^{\rm SEP, eff} = &\, \frac29 a\om_b \la\sqrt{\rhoDM}\sqrt3 \vepS \left[ c_{\ga(t)}\hat{x} - s_{\ga(t)}\hat{y} \right] \,,
\end{align}
while, for \( N=3 \)
\begin{align}
  \vec{F}^{\rm SEP, eff} = &\, -\frac29 a\om_b \la\sqrt{\rhoDM} \vepT \left[ c_{\ga(t)+\chi}\hat{x} + s_{\ga(t)+\chi}\hat{y} \right] \,.
\end{align}
This effect can be used to constrain \( \al \) both for masses near \( \om_b \) (\( N=1 \)) and \( 3\om_b \) (\( N=3 \)) which do not affect secularly \( \dPb \) in the limit \( e\to0 \).  For instance, for the system J1713+0747~\cite{1993ApJ...410L..91F,Zhu:2015mdo,Zhu:2018etc}, the results presented in~\cite{Zhu:2018etc} indicate \( \dot{e}=(-3\pm 4)\times 10^{-18} \text{s}^{-1} \).  Using this and assuming \( N=1 \) with \(   \vepS s_{\ga(t)} \simeq 1 \) and \( N=3 \) with \( \vepT s_{\ga(t)+\chi}\simeq 1 \), we obtain \( \al\lesssim 1.2\times 10^{-2} \) for \( m \simeq 7\times 10^{-22} \)~eV and  \( \al\lesssim 1.8\times 10^{-2} \)  for  \( m \simeq 2\times 10^{-21} \)~eV, respectively, which is competitive with the  current bounds for these masses obtained from the Cassini experiment~\cite{Hohmann:2017uxe}.  We display these constraints in figures~\ref{fig:mono} and~\ref{fig:sin_plus}.

\section{Conclusions}
\label{sec:end}

In this paper we have assessed the constrains that can be achieved on the direct coupling \( \al \) of a spin-2 ULDM field and the stars in binary systems.  As for measurable quantities, we have considered the variation of the orbital parameters of binary pulsars.  Those quantities are secularly affected by the ULDM field when the mass of the field \( m \) is close  to an integer multiple \( N \) of the orbital frequency of the binary system \( \om_b \), i.e., when the resonant condition \( m\simeq N\om_b \) applies.  As studied in~\cite{Blas:2019hxz} for scalar ULDM fields, when detuning from resonance, the effects on the quantities we are considering are suppressed.  For this reason, we have focussed here only on the secular contributions.  We have centered our analysis on the measurements of the secular drift of the orbital period as it is the most constraining one.  However, from the results we provide in Appendix~\ref{app:bessel} it is possible to work out the effects on all orbital parameters.  In particular, as for the scalar field case, there are situations in which the secular variation of \( \dPb \) is negligible, but the secular drift for other parameters is not.  Indeed, we showed cases where the best constraints come from the measurement of \( \dot{e} \).  Furthermore, as emphasised in~\cite{Blas:2019hxz}, it would be worth to perform such analysis to assess whether the constrains obtained either only from \( \dPb \) or only from \( \dot{e} \) can be improved.  In view of the improvement in the precision of future binary pulsar measurements and the increasing number of systems suitable for timing analysis expected for the future from observations as with SKA~\cite{Kramer:2015bea}, from our study we can conclude that constrains up to \( \al\sim10^{-5} \) will be potentially achievable for a considerably large fraction of the range of masses of the ULDM field.

Along the same lines, with the next generation of radio arrays it becomes crucial to take advantage of the large number of systems by developing new statistical approaches and techniques for the extraction of the constraints on the ULDM field.  The main idea is that, by using the whole population of binaries at once, we should be able to boost the signal due to the ULDM field at the expense of the noise.  This is because since the ULDM effects are coherent in time, other secular effects are not expected to be.  We plan to develop several such approaches in a future work.

\acknowledgments

FU wishes to thank E.~Adelberger for useful correspondence, and the Department of Physics, FCEyN UBA, for hospitality while this work was being completed.  FU is supported by the European Regional Development Fund (ESIF/ERDF) and the Czech Ministry of Education, Youth and Sports (MEYS) through Project CoGraDS - \verb|CZ.02.1.01/0.0/0.0/15_003/0000437|.  The work of DNL and JMA has been supported by CONICET, ANPCyT and UBA.  DLN thanks S.~Sibiryakov for useful discussions and comments on the preliminary version of this paper, and to D.~Blas for discussions on related matters.

\appendix

\section{Decomposition of the angular quadrupole matrix}
\label{app:angular}

Following~\cite{Maggiore:1900zz}, we can decompose any symmetric, traceless and constant tensor in terms of spherical harmonics as \( \vep_{ij} \deq \sum_m a_m {\cal Y}^{2m}_{ij} \) where \( Y^{2m} \deq {\cal Y}^{2m}_{ij} n^i n^j \) with \( Y_{2m}(\hat{n}) \) are the real spherical harmonics and \( \hat{n} \deq (x,y,z) \) is the unit coordinate vector (so that \( x^2+y^2+z^2=1 \)).  We normalise the spherical harmonics as
\begin{align}
  & Y^{2,-2} = \sqrt2 xy, &\quad Y^{2,2} = \left(x^2-y^2\right) / \sqrt2 \,, \nn \\
  & Y^{2,-1} = \sqrt2 yz, & \quad Y^{2,1} = \sqrt2 zx \,, \nn \\
  & Y^{2,0} = \left(x^2+y^2-2z^2\right) / \sqrt6 \,.
\end{align}
The multipole matrices then look like
\begin{align}
  & {\cal Y}^{2,-2} = \frac{1}{\sqrt2} \left(
  \begin{array}{ccc}
    0 & 1 & 0 \\
    1 & 0 & 0 \\
    0 & 0 & 0 \\
  \end{array} \right) \,,\quad
  {\cal Y}^{2,2} = \frac{1}{\sqrt2} \left(
  \begin{array}{ccc}
    1 & 0 & 0 \\
    0 & -1 & 0 \\
    0 & 0 & 0 \\
  \end{array} \right) \,, \nn \\
  & {\cal Y}^{2,-1} = \frac{1}{\sqrt2} \left(
  \begin{array}{ccc}
    0 & 0 & 0 \\
    0 & 0 & 1 \\
    0 & 1 & 0 \\
  \end{array} \right) \,,\quad
  {\cal Y}^{2,1} =  \frac{1}{\sqrt2} \left(
  \begin{array}{ccc}
    0 & 0 & 1 \\
    0 & 0 & 0 \\
    1 & 0 & 0 \\
  \end{array} \right) \,, \nn \\
  & {\cal Y}^{2,0} = -\frac{1}{\sqrt6} \left(
  \begin{array}{ccc}
    1 & 0 & 0 \\
    0 & 1 & 0 \\
    0 & 0 & -2 \\
  \end{array} \right) \,,
\end{align}
and the polarisation matrix turns out to be
\begin{align}
  \vep_{ij} & = \frac{1}{\sqrt2}\left(
  \begin{array}{ccc}
    a_{2} - a_{0}/\sqrt3 & a_{-2} & a_{1} \\
    a_{-2} & - a_{2} - a_{0}/\sqrt3 & a_{-1} \\
    a_{1} & a_{-1} & 2a_{0}/\sqrt3 \\
  \end{array} \right) \,.
\end{align}
Our parametrisation \Eq{eq:pol} in terms of tensor, vector, and scalar components can be recovered with the choice
\begin{align}
  & a_{-2} \deq a_\times \deq \vepT \sin\chi \,,\quad a_{2} \deq a_+ \deq \vepT \cos\chi \,, \nn \\
  & a_{-1} \deq a_L \deq \vepV \sin\eta \,,\quad a_{1} \deq a_R \deq \vepV \cos\eta \,, \nn \\
  & a_{0} \deq a_S \deq \vepS \,,
\end{align}
where \( {\vepS}^2+{\vepV}^2+{\vepT}^2=1 \).

An alternative parametrisation in terms of four angles \( (\ze,\be,\xi,\psi) \) can be given as
\begin{align}
  & a_{-2} \deq \sin{\ze}\cos{\be}\sin{\psi} \,,\quad a_{2} \deq \sin{\ze}\cos{\be}\cos{\psi} \,, \nn \\
  & a_{-1} \deq \sin{\ze}\sin{\be}\sin{\xi} \,,\quad a_{1} \deq \sin{\ze}\sin{\be}\cos{\xi} \,, \nn \\
  & a_{0} \deq \cos{\ze} \,.
\end{align}

\section{Elliptic Keplerian orbit and Fourier decomposition}
\label{app:bessel}

We collect here the useful formulas of Keplerian mechanics and the osculating orbits formalism.  More details can be found in~\cite{Danby:1970}.  Following the same notation as in~\cite{Blas:2019hxz,LopezNacir:2018epg} we write down the Lagrange planetary equations,
\begin{subequations}
  \begin{align}
    \frac{\dot{a}}{a}	= &\, \frac{2}{\om_b}\left\{\frac{e\sin\ta}{a\te} F_r + \frac{\te}{r} F_\ta \right\} \,, \label{eq:lag-a}\\
    \dot{e}				= &\, \frac{\te}{a\om_b}\left\{(\cos\ta+\cos E) F_{\ta} + \sin\ta F_{r}\right\} \,, \label{eq:lag-e}\\
    \dot{\Om}			= &\, \frac{r\sin(\ta+\om)}{a^2\om_b\te\sin\io} F_{z} \,, \label{eq:lag-Om}\\
    \dot{\io}			= &\, \frac{r\cos(\ta+\om)}{a^2\om_b\te} F_{z} \,, \label{eq:lag-i}\\
    \dot{\vpi}			= &\, \frac{\te}{ae\om_b}\left\{\left[1+\frac{r}{a\te^2}\right]\sin\ta F_{\ta} - \cos\ta F_{r} \right\} + 2\sin^2\left(\io/2\right)\dot{\Om} \,, \label{eq:lag-vpi}\\
    \dot{\ep_1}			= &\, -\frac{2r}{a^2\om_b} F_{r} + \left(1-\te\right)\dot{\vpi} + 2\te\sin^2\left(\io/2\right)\dot{\Om} \,, \label{eq:lag-ep}
  \end{align}
\end{subequations}
in terms of the following six independent orbital elements: the semimajor axis \( a \), the orbital eccentricity \( e \), the longitude of the ascending node \( \Om \), the longitude of the periastron \( \vpi=\om+\Om \) (with \( \om \) the argument of the periastron, not to be confused with the orbital frequency \(\om_b\)), the time of periastron \( t_0 \), and the inclination angle \( \io \) of the orbital plane with respect to the reference plane of the sky.  Here \( \ep_1=\om_b(t-t_0)+\vpi-\int \dd t \,\om_b \), \( \om_b=\sqrt{GM_T/a^3}=2\pi/P_b \), \( E \) is the eccentric anomaly that is defined by \( \om_b(t-t_0)=E-e \sin E \).  We have also defined \( \te\deq\sqrt{1-e^2} \).  We use cartesian \( (x,y,z) \) and cylindric \( (r,\ta,z) \) coordinates in the orbital plane, and the overdot stands for a derivative with respect to time \( t \).  Therefore, \( \vec{r}\deq\hat{r}=r\cos{\ta}\hat{x}+r\sin\ta\hat{y} \), with \( \ta \) the angular position of \( M_1 \) with respect to the direction of the pericentre, \( \hat{x} \), and we have decomposed the perturbation as \( \vec{F}=F_r\hat{r}+F_{\ta}\hat{\ta}+F_z\hat{z} \).  The expressions of the components of \( \vec{F} \) or a generic vector in the \( (X,Y,Z) \) coordinates can be found in~\cite{Poisson:2014}.

The orbital motion for \( e\neq0 \) can not be expressed in a closed form as a function of time, but can be decomposed as a Fourier series in terms of Bessel functions as:
\begin{subequations}
  \begin{align}
    x/a		= &\, -\frac{3e}{2} + 2\sum \frac{J'_n(ne)}{n} \cos(\nnut) \,, \\
    y/a		= &\, \frac{2\sqrt{1-e^2}}{e} \sum \frac{J_n(ne)}{n} \sin(\nnut) \,, \\
    r/a		= &\, 1 + \frac{e^2}{2}- 2e\sum \frac{J'_n(ne)}{n} \cos(\nnut) \,, \\
    (x/a)^2	= &\, \frac12 + 2e^2 + \sum q_{xx}(ne) \cos(\nnut) \,, \\
    (y/a)^2	= &\, \frac{1-e^2}{2} + \sum q_{yy}(ne) \cos(\nnut) \,, \\
    xy/a^2	= &\, -\frac{8e\sqrt{1-e^2}}{3} + \sum q_{xy}(ne) \sin(\nnut) \,, \\
    (r/a)^2	= &\, 1 + \frac{3e^2}{2} - 4\sum \frac{J_n(ne)}{n^2} \cos(\nnut) \,, \\
    \cos{\ta} = &\, -e + \frac{2(1-e^2)}{e} \sum J_n(ne) \cos(\nnut) \,, \\
    \sin{\ta} = &\, 2\sqrt{1-e^2} \sum J'_n(ne) \sin(\nnut) \,, \\
    (a/r)^2\cos{\ta} = &\, 2\sum n J'_n(ne) \cos(\nnut) \,, \\
    (a/r)^2\sin{\ta}	= &\, \frac{2\sqrt{1-e^2}}{e} \sum n J_n(ne) \sin(\nnut) \,, \\
    \sin^2{\ta} = &\, B_0 +\sum B_n(ne) \cos(\nnut) \,, \\
    \sin{\ta}\cos{\ta} = &\, \tilde{B}_0+\sum \tilde{B}_n(ne)\sin(\nnut) \,, 
  \end{align}
\end{subequations}
where the sums run over \( n\in[1,\infty) \).  The expansion coefficients are
\begin{subequations}
  \begin{align}
    n q_{xx}(ne)	\deq &\, J_{n-2}(ne) - J_{n+2}(ne) - 2e\left[ J_{n-1}(ne) - J_{n+1}(ne) \right] \\
				= &\, 4J_{n}' (ne)\frac{(1-e^2)}{e}-\frac{4 J_n(ne)}{n e^2} \,,\nonumber \\
    n q_{yy}(ne)	\deq &\, (1-e^2)\left[ J_{n+2}(ne) - J_{n-2}(ne) \right] \\
				= &\, - n q_{xx}(ne) - 4J_n(ne)/n \,, \nonumber\\
    n q_{xy}(ne)	\deq &\, \sqrt{1-e^2}\left[ - 2J_n(ne) + J_{n+2}(ne) + J_{n-2}(ne) \right] \\
				= &\, 4\sqrt{1-e^2}\left[ \, J_{n}(ne)\frac{(1-e^2)}{e^2} - \frac{J_n'(ne)}{n e} \right] \,, \nonumber
  \end{align}
\end{subequations}
where the \( J_n(z) \) are Bessel functions of the first kind.  The coefficients \( B_n(ne) \) and \( \tilde{B}_n(ne) \) are more complex, and can be written in terms of series of Bessel functions as we show next (see also~\cite{Watson1995treatise}).

\paragraph{Expansion of \( \sin^2{\ta} \)}
\begin{align}
  \sin^2{\ta} &\, = \frac{\te^2\sin^2{E}}{(1-e\cos{E})^2} \deq B_0+\sum_n B_n \cos{(n\om_b t)} \,,
\end{align}
where as usual \( E \) is the \textit{eccentric anomaly} defined by \( \om_b t \deq E-e\sin{E} \).
\begin{align}
  B_0 = &\, \frac{\te^2}{1+\te} \,, \nn \\
  B_n = &\, \frac{\te}{2} \left[ 2J_n-J_{n+2}-J_{n-2} \phantom{ + \sum_q} \right. \nn \\
    + &\, \left. \sum_q \mathcal{E}^q \left( 2J_{n+q}-J_{n+q+2}-J_{n+q-2} + 2J_{n-q}-J_{n-q+2}-J_{n-q-2} \right) \right] \,. \label{Bn}
\end{align}
 where the sums run over \( q \in [1,+\infty) \) and \( \mathcal{E} \deq e/(1+\te) \).

\paragraph{Expansion of \( \sin{\ta}\cos{\ta} \)}
\begin{align}
  \sin{\ta}\cos{\ta} &\, = \frac{\te\sin{E}(\cos{E}-e)}{(1-e\cos{E})^2} \deq \tilde{B}_0+\sum_n \tilde{B}_n \sin{(n\om_b t)} \,,
\end{align}
where
\begin{align}
  \tilde{B}_0 = &\, \frac{\te}{\pi e^2}\left[ \te^2\log\left(\frac{1+e}{1-e}\right)-2e \right] \,, \nn \\
  \tilde{B}_n = &\, J'_{n+1}+J'_{n-1}-2eJ'_n \nn \\
    + &\, \sum_q \mathcal{E}^q \left( J'_{n+q+1}+J'_{n+q-1}-2eJ'_{n+q}+J'_{n-q+1}+J'_{n-q-1}-2eJ'_{n-q} \right) \,. \label{Bntilde}
\end{align}

For computational reasons we can not run the \( q \) sum to infinity, so we have to limit ourselves to a finite number of terms.  We have studied the convergence of the \Eqs{Bn}{Bntilde} series numerically and we have found that, at least out to the harmonics of interest for us, that is, \( N=5 \), truncating the sum at \( q=3(n+2) \) (where \( n \) is the harmonic we are computing) the series converges rapidly and the result approximates the full series to better than percent level.

Using the expansion of the Bessel function, $ J_N(x)\simeq (x/2)^N/N!$, and its derivative for small values of \( e \), up to corrections of order \( e^2\),we get 
\begin{subequations}\label{Circularlimit}
  \begin{align}
    & e J'_N(Ne) \simeq J_N(Ne) \to  \frac{e}{2}\delta_{N,1} \,, \\
    &\tilde{B}_N(Ne) \simeq -{B}_N(Ne) \to \frac{1}{2}\delta_{N,2}+e (\delta_{N,3}-\delta_{N,1}) \,.
  \end{align}
\end{subequations}

In Panel~\ref{orb.param} we collect the six orbital parameters written in terms of the polarisation components of \( \vep_{ij} \); then, in Panel~\ref{orb.param.Bessel}, we show the explicit expressions for the secular changes of all orbital parameters in terms of Bessel functions.

\begin{panel*}[htbp]
\caption{Expressions for all six orbital parameters in terms of the polarisations.  We use the shortcut notation \( s_x \deq \sin{x} \), \( c_x \deq \cos{x} \).}
\begin{subequations}
  \begin{align}
    \frac{\dot{a}}{a} = &\, \frac{4\la\sqrt{\rhoDM}}{\te} \left\{ \wm \left(\frac{a}{r}\right)^2 \left[ \vepT (s_{\chi-2\ta} + e  s_{\chi-\ta}) - e\frac{\vepS}{\sqrt3} s_\ta \right] c_{mt+\Upsilon} \right. \nn \\
    &\, \left. -\frac{1}{\te} \left[ \vepT(c_{\chi-2\ta} + 2e   c_{\chi-\ta} +  e^2   c_\chi )+ \frac{\vepS}{\sqrt3}(1+e^2+2e c_\ta) \right] s_{mt+\Upsilon} \right\} \\
    \dot e = &\, \frac{2\la\te\sqrt{\rhoDM}}{e} \left\{ \wm \left(\frac{a}{r}\right)^2 \left[ \left(1-\frac{r}{a}\right) \vepT s_{\chi-2\ta} + e \vepT s_{\chi-\ta} - e\frac{\vepS}{\sqrt3} s_\ta \right] c_{(mt+\Upsilon)} \right. \nn \\
    &\, -\frac{1}{\te} \left[ \left(1-\frac{r}{a}\right) \vepT c_{\chi-2\ta} + \left(2-\frac{r}{a}\right) e \vepT c_{\chi-\ta} + 2\frac{\vepS}{\sqrt{3}} e(e+ c_\ta) \left.  + e^2 \vepT c_\chi \right] s_{mt+\Upsilon} \right\} \\
    \dot\Om = &\, \frac{2\la\sqrt{\rhoDM}\vepV}{\te\sin{\io}} \left\{ \wm \left(\frac{a}{r}\right)  c_{\eta-\ta} s_{\ta+\om} c_{mt+\Upsilon} + \frac{1}{\te}\left(\frac{r}{a}\right)  \left(  s_{\eta-\ta} + e s_\eta\right) s_{\om+\ta} s_{mt+\Upsilon}\right\} \\
    \dot\io = &\, \frac{2\la\sqrt{\rhoDM}\vepV}{\te} \left\{ \wm \left(\frac{a}{r}\right) c_{\eta-\ta}    c_{\ta+\om} c_{mt+\Upsilon}  + \frac{1}{\te}\left(\frac{r}{a}\right)   c_{\ta-\om} \left(s_{\eta-\ta}+ e s_\eta \right) s_{mt+\Upsilon}\right\} \\
    \dot\vpi = &\, \frac{2\la\te\sqrt{\rhoDM}}{e^2} \left\{ \wm \left(\frac{a}{r}\right)^2 \left[ \vepT c_{\chi-2\ta} - e \vepT c_{\chi-\ta} + e\frac{\vepS}{\sqrt3} c_\ta \right] c_{mt+\Upsilon} \right. \nn \\
    &\, -\frac{1}{\te^2} \wm \left(\frac{a}{r}\right) \vepT \left( c_{\chi-2\ta} + e   c_{\chi-\ta} \right)c_{mt+\Upsilon} -\frac{1}{\te} \left[ \vepT (s_{2\ta-\chi}+ e^2  s_\chi)+2 e\frac{\vepS}{\sqrt3} s_\ta\right] s_{mt+\Upsilon} \nn \\
    &\, \left. -\frac{1}{\te^3} \left(\frac{r}{a}\right) \vepT \left( s_{\chi-2\ta} + 2e   s_{\chi-\ta} +  e^2   s_\chi \right) s_{mt+\Upsilon} \right\} + 2s^2_{\io/2}\dot{\Om} \\
    \dot\ep_1 = &\, -4\la\sqrt{\rhoDM} \left\{ \wm \left(\frac{a}{r}\right) \left[ \vepT c_{\chi-2\ta} - \frac{\vepS}{\sqrt3} \right] c_{mt+\Upsilon} \right. \nn \\
    &\, \left. +\frac{1}{\te} \left(\frac{r}{a}\right) \left[ \vepT s_{\chi-2\ta} + e \vepT s_{\chi-\ta} - e\frac{\vepS}{\sqrt3} s_\ta \right] s_{mt+\Upsilon} \right\} + \left(1-\te\right)\dot{\vpi} + 2\te s^2_{\io/2}\dot{\Om}
  \end{align}
\end{subequations}\\
\centering\rule[12pt]{0.485\textwidth}{0.1pt}
\label{orb.param}
\end{panel*}

\begin{panel*}[htbp]
\caption{Secular changes of all six orbital parameters.}
\begin{subequations}
  \begin{align}
    \avg{\frac{\dot{a}}{a}}	= &\,  -\frac43 \la \sqrt{3\rhoDM}\bigg\{ \vepS J_N(Ne)s_{\ga(t)} + \vepT\bigg[\mathcal{F}_+(N,e)s_{\ga(t)+\chi}+\mathcal{F}_-(N,e)s_{\ga(t)-\chi} \bigg] \bigg\} \\ 
   \nonumber &\, \\
    \nonumber \avg{\dot e}          	= &\, -\frac23\la\sqrt{3\rhoDM}\frac{\te^2}{eN}\bigg\{ \vepT\bigg[\mathcal{F}_+(N,e)\bigg(N-\frac{2}{\te} \bigg)s_{\ga(t)+\chi}+\mathcal{F}_-(N,e)\bigg(N+\frac{2}{\te} \bigg)s_{\ga(t)-\chi} \bigg] \\
    &\, + \vepS N J_N(Ne)s_{\ga(t)} \bigg\}   \\
    \nonumber &\, \\
    \nonumber \avg{\dot\Om}				= &\,  -\frac{\la\sqrt{\rhoDM}}{Ns_\io}\vepV\bigg\{ c_{\ga(t)}\bigg[ \bigg(\frac{B_N(Ne)}{\te^3}-\frac{2J_N(Ne)}{\te}\bigg)s_{\eta-\om}-\frac{2J_N(Ne)}{\te}s_{\eta+\om} \bigg]\\
    &\,  - s_{\ga(t)}\bigg[\frac{\tilde{B}_N(Ne)}{\te^3}+\frac{2eJ'_N(Ne)}{\te^2} \bigg]c_{\eta-\om}\bigg\} \\
    \nonumber &\, \\
    \nonumber \avg{\dot\io}	           = &\, \frac{\la\sqrt{\rhoDM}}{N}\vepV\bigg\{ c_{\ga(t)}\bigg[\bigg(\frac{B_N(Ne)}{\te^3}-\frac{2J_N(Ne)}{\te}\bigg)c_{\eta-\om}+ \frac{2J_N(Ne)}{\te}c_{\eta+\om}\bigg]\\
    &\,  + s_{\ga(t)}\bigg[\frac{\tilde{B}_N(Ne)}{\te^3}+\frac{2eJ'_N(Ne)}{\te^2}\bigg]s_{\eta-\om}\bigg\}\\
    \nonumber &\, \\
    \nonumber \avg{\dot\vpi}			= &\, \frac23\la\sqrt{3\rhoDM}\frac{\te^2}{e^2N}\bigg\{  \vepT\bigg[c_{\ga(t)-\chi}\bigg( \mathcal{F}_-(N,e)\bigg( N+\frac{2-e^2}{\te^3} \bigg)+\frac{\sqrt{3}e^2J_N(Ne)}{2\te^3}\bigg) \\
    \nonumber &\,-c_{\ga(t)+\chi}\bigg( \mathcal{F}_+(N,e)\bigg( N-\frac{2-e^2}{\te^3}\bigg)-\frac{\sqrt{3}e^2J_N(Ne)}{2\te^3}\bigg)\bigg]  -\frac{\vepS}{\te} N e  J'_N(Ne) c_{\ga(t)} \bigg\} \\
    &\,  + 2s^2_{\io/2}\avg{\dot{\Om}} \\
    \nonumber &\, \\
    \nonumber \avg{\dot\ep_1}			= &\,\frac83\frac{\la\sqrt{3\rhoDM}}{N}\bigg\{ \vepT \bigg[ c_{\ga(t)+\chi}\mathcal{F}_+(N,e)+c_{\ga(t)-\chi}\mathcal{F}_-(N,e)\bigg] + \vepS J_N(Ne)c_{\ga(t)}\bigg\}\\
    &\, + (1-\te)\avg{\dot\vpi} + 2\te s^2_{\io/2}\avg{\dot{\Om}}
  \end{align}
\end{subequations}\\
\centering\rule[12pt]{0.485\textwidth}{0.1pt}
\label{orb.param.Bessel}
\end{panel*}

\section{Data}
\label{app:pulsars}

Table~\ref{tab:frac} lists all the binary systems that we have used in this study, alongside their relevant properties.

\begin{table*}[htbp]
  \centering
  \begin{tabular}[h]{lccccccr}
    \toprule
    Name & \(M_1\) [\(M_\odot\)] & \(M_2\) [\(M_\odot\)] & \(e\) & \(P_b\) & \(\dPb^\text{int}\) & \(\delta\dPb^\text{int}\) & References \tabularnewline
    \midrule
    J1903+0327	& 1.0		& 1.7		& 0.44		& 95	& -6.4e-11		& 3.1e-11		& \cite{Freire:2010tf} \tabularnewline
	J1740-3052	& 20		& 1.4~\P	& 0.58		& 231	& NA			& 3.0e-9~\dag	& \cite{Madsen:2012rs} \tabularnewline
	J0737-3039	& 1.2		& 1.3		& 0.088		& 0.10	& -4.0e-15		& 1.7e-14		& \cite{Kramer:2006nb,Wex:2014nva} \tabularnewline
	B1913+16	& 1.4		& 1.4		& 0.62		& 0.32	& 5.0e-15		& 4.0e-15		& \cite{Weisberg:2016jye} \tabularnewline
	B1259-63	& 24		& 1.4~\P	& 0.87		& 1237	& 1.0e-9		& 7.0e-9		& \cite{Shannon:2013dpa} \tabularnewline
	J1012+5307	& 0.16		& 1.6		& 1.3e-6	& 0.60	& -1.8e-14		& 2.1e-14		& \cite{1998MNRAS.298..207C,Desvignes:2016yex,Arzoumanian:2017puf} \tabularnewline
	J1614-2230	& 0.49		& 1.9		& 1.3e-6	& 8.7	& 3.4e-13		& 2.0e-13		& \cite{Fonseca:2016tux,Arzoumanian:2017puf} \tabularnewline
	J1909-3744	& 0.21		& 1.5		& 1.2e-7	& 1.5	& -4.0e-15		& 1.4e-14		& \cite{Desvignes:2016yex,Arzoumanian:2017puf} \tabularnewline
	J0751+1807	& 0.16		& 1.6		& 3.3e-6	& 0.26	& -4.6e-14		& 3.5e-15		& \cite{Desvignes:2016yex} \tabularnewline
	J1910+1256	& 0.19~\S	& 1.6		& 2.3e-4	& 58	& -2.0e-11		& 4.0e-11		& \cite{Gonzalez:2011kt} \tabularnewline
	J2016+1948	& 0.29~\S	& 1.0		& 1.5e-3	& 635	& -1.0e-9		& 2.0e-9		& \cite{Gonzalez:2011kt} \tabularnewline
	J0348+0432	& 0.17		& 2.0		& 2.4e-6	& 0.10	& -1.1e-14		& 4.5e-14		& \cite{Antoniadis:2013pzd} \tabularnewline
	J1713+0747	& 0.29		& 1.3		& 7.5e-5	& 68	& 3.0e-14		& 1.5e-13		& \cite{Zhu:2018etc} \tabularnewline
	J0613-0200	& 0.12~\S	& 1.2~\P	& 5.4e-6	& 1.2	& 2.7e-14		& 1.0e-14		& \cite{Desvignes:2016yex,Arzoumanian:2017puf} \tabularnewline
	J1738+0333	& 0.19		& 1.5		& 3.4e-7	& 0.35	& 2.0e-15		& 4.0e-15		& \cite{Freire:2012mg} \tabularnewline
	J1751-2857	& 0.18~\S	& 1.2~\P	& 1.3e-4	& 111	& NA			& 1.8e-11~\dag	& \cite{Desvignes:2016yex,Caputo:2017zqh} \tabularnewline
	J1857+0943	& 0.24		& 1.4		& 2.2e-4	& 12	& NA			& 1.2e-13~\dag	& \cite{Desvignes:2016yex,Caputo:2017zqh,Arzoumanian:2017puf} \tabularnewline
    \bottomrule
  \end{tabular}
  \caption{List of binary systems used in this study.  The columns are: (1) the name of the binary; (2) the mass of the companion in \( M_\odot \) units -- if only the minimum value is available we denote this with a \S; (3) the mass of the pulsar in \( M_\odot \) units -- assumed values are indicated with a \P; (4) the orbital eccentricity; (5) the binary period in days; (6) the ``intrinsic'' period derivative in \(\text{s}\,\text{s}^{-1}\) (see the text for our definition of ``intrinsic'') -- ``NA'' means that only an upper limit on the measured \(\dPb^\text{obs}\) value was given, which we report as the error in the next column; (7) the error on the ``intrinsic'' period derivative, also in \( \text{s}\,\text{s}^{-1} \) -- the \dag~indicates an upper limit; (8) the references.}\label{tab:frac}
\end{table*}

\clearpage
\bibliographystyle{hieeetr}
\bibliography{biblio.bib}

\end{document}